\documentclass{aa}  

\usepackage{graphicx}
\usepackage{txfonts}
\usepackage{lipsum}
\usepackage{textgreek}
\usepackage{subcaption}         
\usepackage{lscape}             
\usepackage{placeins}           
\usepackage{ulem}
\usepackage[colorlinks=true, citecolor=blue]{hyperref}

\begin{document}

   \title{A precessing AGN with S-shaped morphology discovered in MALS}

   \author{J.D. Wagenveld\inst{1,2}
           \and N. Gupta\inst{3}
        }

   \institute{Max-Planck-Institut f\"ur Radioastronomie, 
              Auf dem H\"ugel 69, 53121 Bonn, Germany
        \and Department of Astrophysics, University of Oxford, 
             Denys Wilkinson Building, Keble Road, 
             Oxford OX1 3RH, UK
        \and Inter-University Centre for Astronomy and Astrophysics,
             Post Bag 4, Ganeshkhind, 
             Pune 411 007, India}

   \date{Received September 30, 2026}

 
  \abstract{
  We present J1221$-$1354, a precessing AGN discovered in the MeerKAT Absorption Line Survey (MALS) using deep L-band (0.9--1.6\,GHz) radio data. Its `S'-shaped morphology, revealed by the excellent surface brightness sensitivity of MeerKAT, is consistent with jet precession. Ancillary optical and infrared data establish the source as a Type 2 AGN in a moderately overdense environment. 
  The approaching (southern) jet is brighter but shorter than the receding jet, with the northern edge of the receding jet showing a significantly younger spectral age. These features are peculiar, but consistent with a jet reorientation event in an asymmetric ambient medium, in which the jet re-enters a cavity of previously deposited plasma. A simple ballistic model yields a precession period of $P=290\pm30$\,Myr, consistent with geodetic precession in a supermassive black hole binary (SMBHB) with a binary separation of less than 1.1\,pc. We expect that deep radio continuum surveys in the near future will uncover many more such systems, helping to further strengthen the connection between jet precession and SMBHBs, and the latter's implications for the gravitational-wave background detected with pulsar timing arrays.
  }


   \maketitle
   \nolinenumbers

\section{Introduction}

The study of active galactic nuclei (AGN) yields important insights into the formation and evolution of galaxies and their central black holes. At radio wavelengths, many AGN show powerful jets; while most radio galaxies show symmetric opposing jets, peculiar morphologies are also common, arising from jet interaction with the surrounding medium, most prominently in galaxy clusters, or from jet precession. Precessing jets are identified by their inversion-symmetric morphology, which, depending on precession angle and viewing inclination, ranges from subtle jet bending to `S'- and `Z'-shaped radio galaxies \citep[e.g.,][]{Nolting2023}. A number of these were already discovered in early radio surveys \citep[e.g.,][]{Gower1982,Condon1984a,Hunstead1984,Taylor1990}. Though the exact cause of precession cannot always be determined, geodetic precession of the black hole spin axis is predicted to arise in a supermassive black hole binary \citep[SMBHB,][]{Begelman1980,Roos1988}, making jet precession a possible indicator of a SMBHB. Indeed, for a number of precessing AGN, the precession is consistent with that induced by a binary companion to the central black hole \citep{Roos1988,Krause2019}, or the orbital motion of the binary itself \citep{Roos1993}. 

Modern radio interferometers with a large number of antennas concentrated in a central core, such as MeerKAT \citep{Jonas2016}, the LOw-Frequency ARray \citep[LOFAR,][]{vanHaarlem2013}, and the Australian Square Kilometre Array Pathfinder \citep[ASKAP,][]{Hotan2021}, offer excellent sensitivity to faint diffuse radio emission. Surveys with these telescopes are already providing important insights into the physics behind a variety of jet morphologies. A well-known example is the `X'-shaped radio galaxy PKS\,2014--55, where MeerKAT data indicated a clear preference for hydrodynamical backflow over jet reorientation \citep{Cotton2020}. Other studies have revealed the influence of the cluster environment \citep[e.g.][]{Chibueze2021}, jet precession \citep[e.g.][]{Nandi2021,Sethi2024,Misra2025}, and in some cases, both simultaneously \citep[e.g.][]{Koribalski2024,Kumari2024}. Ancillary optical and radio data, in particular very long baseline interferometry (VLBI), have revealed that some AGN with precessing jets show double-peaked optical emission lines and multiple compact components associated with the core at milli-arcsecond (mas) scales \citep{Nandi2021}. These present strong arguments for the SMBHB origin, however, definitive proof remains elusive. 

\begin{figure*}
    \centering
    \includegraphics[width=0.48\linewidth]{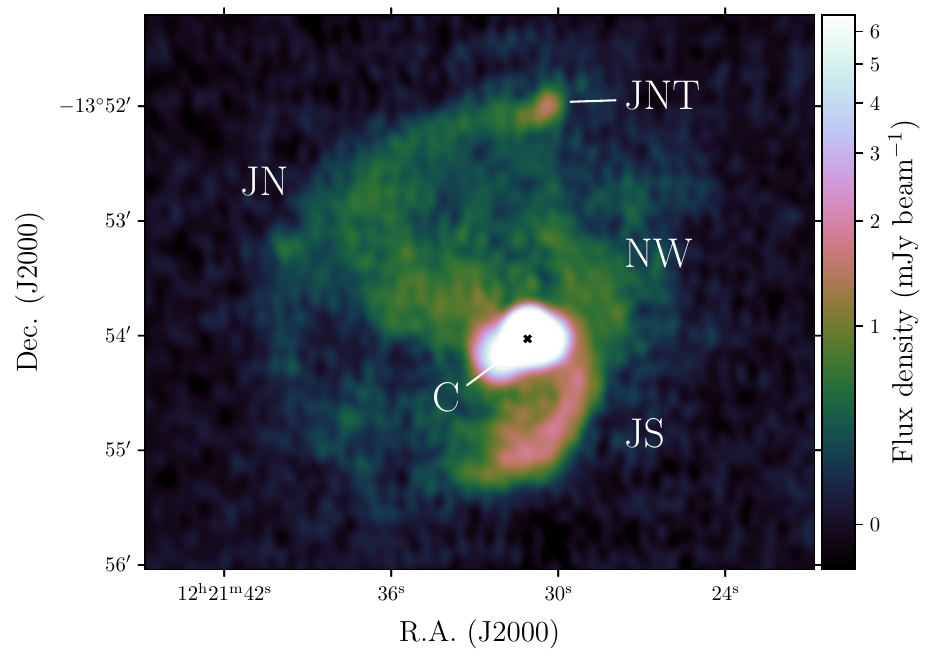}
    \includegraphics[width=0.48\linewidth]{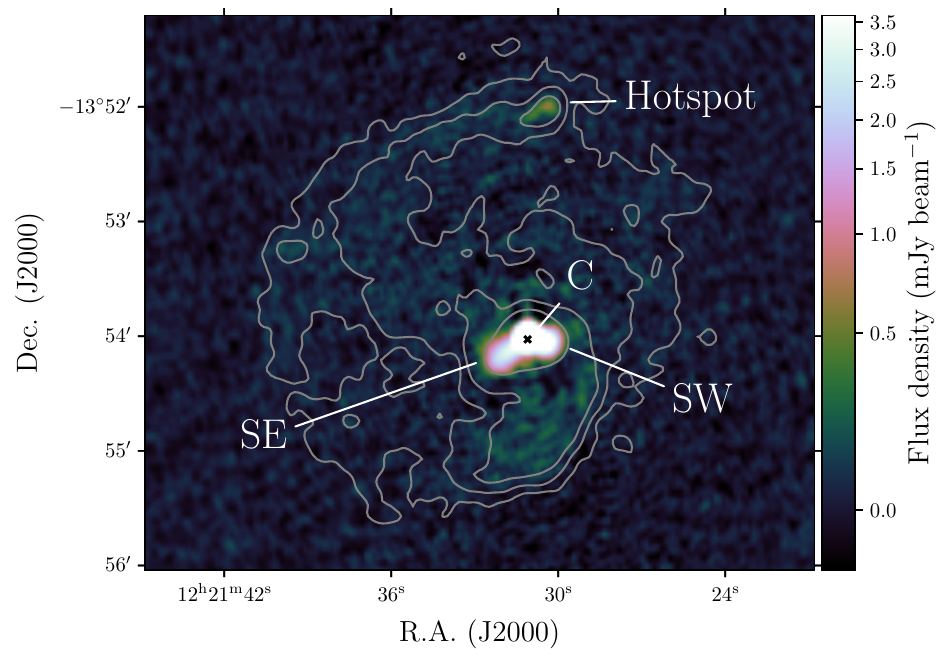}
    \includegraphics[width=0.48\hsize]{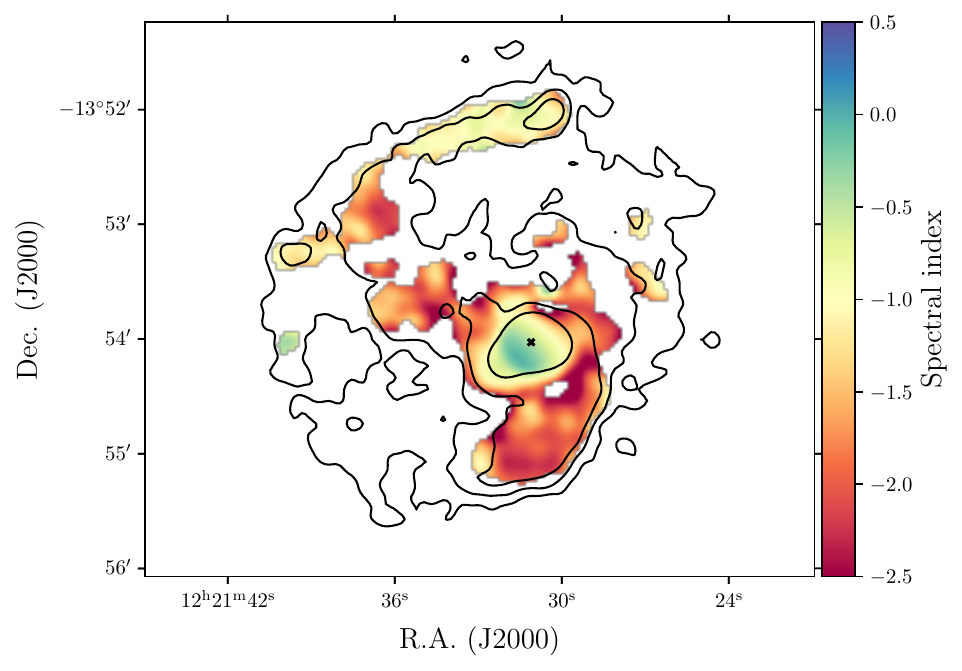}
    \includegraphics[width=0.48\hsize]{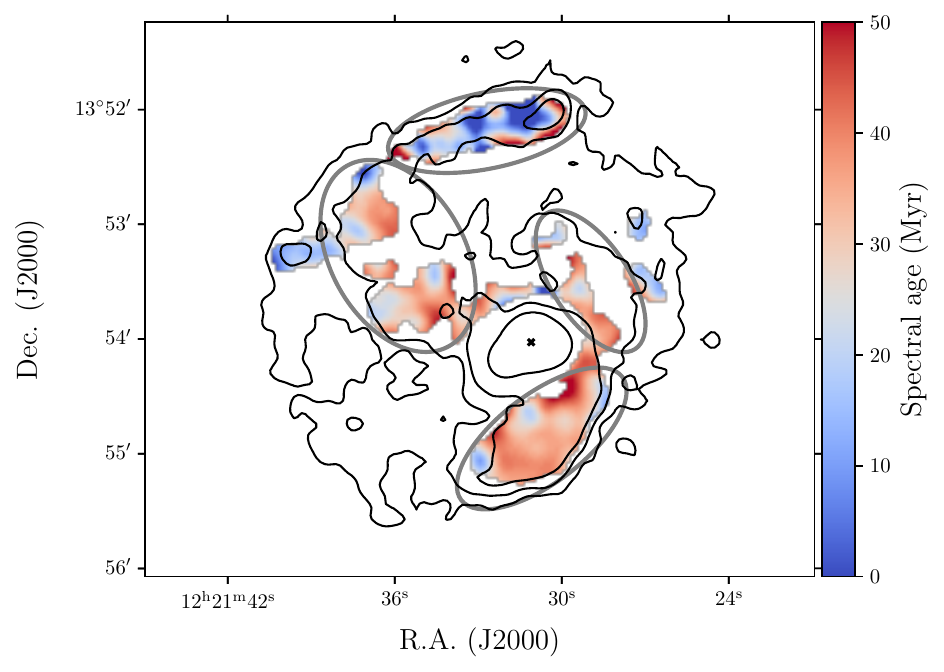}
    \caption{Radio continuum properties of J1221$-$1354 as seen by MeerKAT. \textit{Top-left:} 0.9\,GHz intensity ($12\farcs3\times11\farcs0$ beam), with the core (C), southern jet (JS), northern jet (JN), north-west (NW) component, and northern jet tip (JNT) labelled. \textit{Top-right:} 1.6\,GHz intensity ($6\farcs8\times6\farcs2$ beam), with the south-east (SE) and south-west (SW) components labelled. \textit{Bottom-left:} Spectral index over 0.9--1.6\,GHz (see Sect.~\ref{sec:specage}). \textit{Bottom-right:} Spectral age derived from spectral fitting with BRATS (see Sect.~\ref{sec:specage}). The individual regions used for the analysis are demarcated by grey ellipses (see also Table~\ref{tab:specage}). Spectral ages for JNT have been fit with separately with a different magnetic field strength.
    In all panels, contours correspond to 0.9\,GHz emission at 0.2, 0.5, 1.0 and 5\,mJy\,beam$^{-1}$, and the cross marks the optical position of the host galaxy.
    }
    \label{fig:radio_continuum}
\end{figure*}

In this paper, we present the discovery and analysis of a likely precessing AGN, J1221$-$1354, using images from the MeerKAT Absorption Line Survey \citep[MALS,][]{Gupta2016}. While the source has been observed in other surveys, the excellent surface brightness sensitivity and resolution of MeerKAT allow us, for the first time, to reveal and characterise its precessing jet morphology. We describe the MeerKAT and ancillary data in Sect.~\ref{sec:data}, present results on the host galaxy, radio morphology, and environment in Sect.~\ref{sec:results}, discuss their implications in Sect.~\ref{sec:discussion}, and conclude in Sect.~\ref{sec:conclusion}. Throughout the paper, we use a Planck 2018 cosmology \citep{PlanckCollaboration2018}, with $H_0=67.7\ \mathrm{km\ s}^{-1}\ \mathrm{Mpc}^{-1}$, $\Omega_m=0.31$, and $\Omega_{\Lambda}=0.69$. In this cosmology, at $z$ = 0.1027, 1$\arcsec$ corresponds to 1.95\,kpc. For spectral index ($\alpha$), we use the convention $S\propto\nu^\alpha$, where $S$ is the flux density and $\nu$ the frequency.

\section{Data}\label{sec:data}

MALS observed $\sim$400 pointings centred on bright ($\gtrsim$200\,mJy) AGN, yielding high-fidelity continuum images and catalogues across 15 spectral windows (SPW0--14) of L-band spanning 900--1670\,MHz, with each SPW covering approximately 60 MHz \citep[DR1,][]{Deka2024a}, and full-bandwidth images \citep[DR2,][]{Wagenveld2024a}. Here we use MALS DR1 products for morphological and spectral analysis of J1221$-$1354, the central target of pointing J122131.10-135402.3. The H~{\sc i} 21-cm line, OH 18-cm main and satellite (1720\,MHz) lines corresponding to the source redshift ($z=0.1027$) are all affected by radio frequency interference (RFI), precluding any absorption line search \citep[see][their Fig.~2]{Gupta2021}. We complement the MeerKAT continuum data with Very Large Array (VLA) Sky Survey \citep[VLASS,][]{Lacy2020} image cutouts ($\sim$3\,GHz) and VLBI data \citep{Petrov2025} at higher resolution, the 6dF Galaxy Survey \citep[6dFGS,][]{Jones2004} optical spectrum, the Dark Energy Spectroscopic Instrument (DESI) Legacy Surveys \citep{Dey2019a} optical images, and infrared spectra from the Spectro-Photometer for the History of the Universe, Epoch of Reionization, and Ices Explorer \citep[SPHEREx,][]{Bock2026}. 

\section{Results}\label{sec:results}

\begin{figure*}
    \centering
    \hbox{
    \includegraphics[width=0.55\hsize]{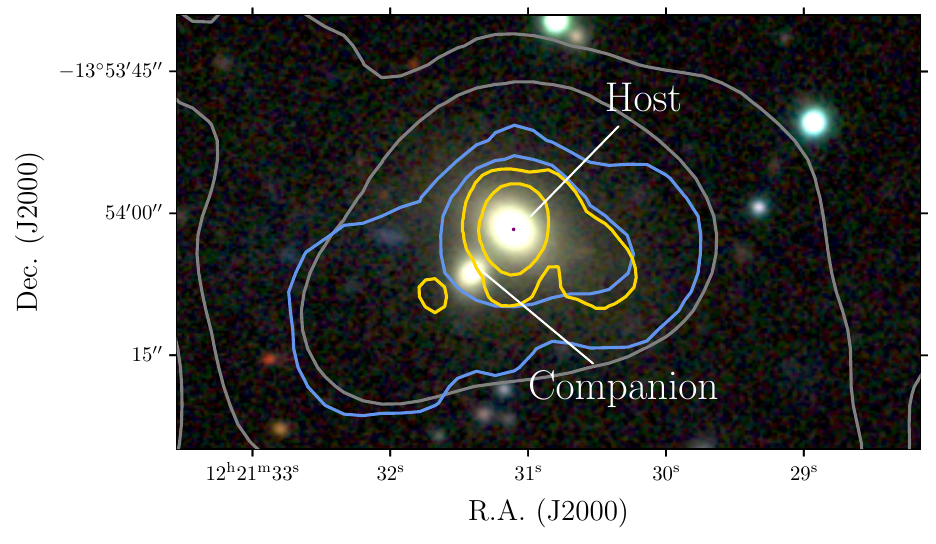}
    \includegraphics[width=0.4\hsize]{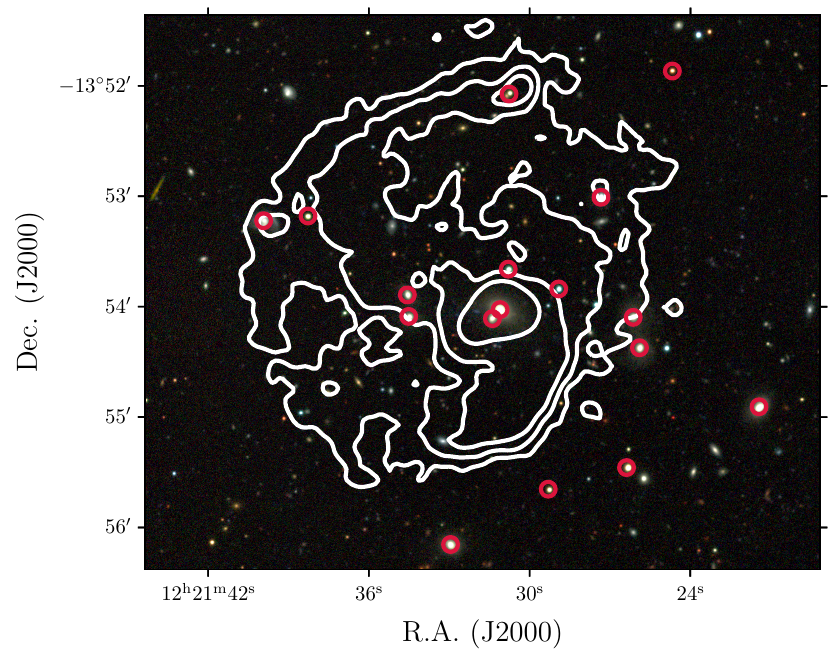}
    }
    \caption{DESI $z$,$r$,$g$ filter composite image of J1221$-$1354 at two scales, overlaid with the radio contours.
    \textit{Left:} Zoomed-in view with the host galaxy and potential companion labelled. Radio contours from MALS at 0.9\,GHz (grey; 0.2, 0.5, 1.0, and 5.0 mJy\,beam$^{-1}$), 1.6\,GHz (blue; 0.5, 5.0 mJy\,beam$^{-1}$), and VLASS at 3\,GHz (yellow; 0.6, 5.0 mJy\,beam$^{-1}$) are overlaid. The purple dot marks the compact VLBI core. 
    \textit{Right:} Large-scale environment with MALS 0.9\,GHz contours at the same levels as the left panel. Circles mark sources for which SPHEREx spectra were fitted to obtain redshifts.
    }
    \label{fig:optical}
\end{figure*}

\subsection{Radio morphology}
\label{sec:radio}

The MeerKAT radio maps of J1221$-$1354 at 0.9\,GHz ($12\farcs3\times11\farcs0$ beam; SPW0) and 1.6\,GHz ($6\farcs8\times6\farcs2$ beam; SPW14) from MALS DR1 are shown in the top row of Fig.~\ref{fig:radio_continuum}. The radio morphology comprises a bright core `C' coinciding with the host galaxy, and extended jet emission, JS and JN, arranged in a spiral `S' shape, clearly visible at lower frequencies (\textit{top-left} panel). Beyond these jets, more diffuse emission is also visible, including a diffuse component north-west (NW) of the core. At 0.9\,GHz, JS is significantly brighter than JN. The jets disappear almost completely at 1.6\,GHz (\textit{top-right} panel), indicating strong spectral evolution. At the higher resolution of the 1.6\,GHz image, two components south-east (SE) and south-west (SW) are visible on either side of C. SW is consistent with the expected direction of JS close to the core. SE is at a right angle to the main jet axis and therefore inconsistent with a counter-jet (details in Section~\ref{sec:host}). SW is solidly detected in the VLASS image while SE is tentative (Fig.~\ref{fig:optical}).
Though these components are not resolved at the lower MALS frequencies, their VLASS images at 3\,GHz allow for a characterisation of their spectrum. Comparing peak surface brightness of the SW and SE components between VLASS and MALS at 1.6 GHz, we find $\alpha=-2.3$ for SW, and $\alpha<-1.9$ for SE given that the VLASS detection is tentative. These values are consistent with those of the other jet components (see Tab.~\ref{tab:specage}), suggesting that these components are indeed also part of the larger jet system. For the core C, $\alpha=-0.6$ between MALS 1.6 GHz and VLASS, consistent with the value determined from the MALS data alone.
At the highest resolutions, VLBI data show a prominent compact component at both C- (4.3\,GHz, 8.5\,mas\,$\times$\,2.1\,mas beam) and X- (7.6\,GHz, 4.6\,mas\,$\times$\,1.2\,mas beam) bands. Single-component fits yield flux densities of $49\pm2$\,mJy and $39\pm2$\,mJy at C- and X-bands, respectively, consistent with a flat spectrum. At X-band, another faint component is visible north-east of C. The nature of this component is discussed further in Sect.~\ref{sec:discussion}.

\subsection{Host galaxy properties and environment}
\label{sec:host}

The host galaxy appears as an elliptical in DESI images, bright in both optical and infrared. As seen in Fig.~\ref{fig:optical} (left), it shows a smooth, regular morphology with no tidal features, shells, or asymmetry, suggesting that any merger occurred in the distant past. Its WISE colours, $W1-W2 = 0.21$ and $W2-W3 = 1.35$, are as expected consistent with an elliptical galaxy classification \citep{Wright2010}. The host has been observed by 6dFGS \citep[][]{Jones2004}, providing a redshift of $z=0.1027$, measured from stellar absorption features \citep[][]{Jones2009}.  
The observed spectrum presented in Appendix~\ref{app:6dfspec} shows several prominent emission lines. The spectrum also shows low-level continuum undulations broader than the emission lines of interest.  We fit each emission line locally, with H$\alpha$ and  [\ion{O}{iii}] independently yielding consistent widths (FWHM $\approx$ 550\,km\,s$^{-1}$) and no evidence for an additional broad or velocity-offset component. The narrow width of H$\beta$ is attributed to unmodelled stellar absorption (see App.~\ref{app:6dfspec}).

The analysis confirms the source as a Type 2 AGN based on the Baldwin-Phillips-Terlevich \citep[BPT,][]{Baldwin1981} line ratios, $\log([\ion{O}{iii}]/\mathrm{H}\beta) \leq 0.89$ and $\log([\ion{N}{ii}]/\mathrm{H}\alpha) = -0.23 \pm 0.09$, consistent with the classification of \citet[][]{Suresh2025}.   

In the optical DESI image (Fig.~\ref{fig:optical}; \textit{right}), a number of nearby galaxies stand out due to their similarity in colour and galaxy type to the host. Using SPHEREx spectra of nearby bright sources within $2\farcm5$ of J1221$-$1354 (see App.~\ref{app:spherex}), we find that seven sources have redshifts consistent with the host redshift, suggesting a moderately over-dense environment. 
The SPHEREx redshift of J1221$-$1354 itself is consistent with the spectroscopic redshift within 3$\sigma$, validating the method. The search radius of $2\farcm5$ ($\sim$290\,kpc at $z$ = 0.1027) is sufficient to cover the full extent of the radio emission and sample the immediate environment of the host. The closest galaxy (6$\arcsec$; `companion' in Fig.~\ref{fig:optical}) is identified as a foreground source at $z$ = 0.058 $\pm$ 0.002, though its proximity to J1221$-$1354 means its SPHEREx spectrum may be contaminated. Notably, the direction of the SE (Fig.~\ref{fig:radio_continuum}, \textit{top-right} panel) is consistent with the direction of this `companion' galaxy, though it extends far beyond its optical radius. 
%

\subsection{Spectral modelling}
\label{sec:specage}

\begin{table*}[]
    \centering
    \renewcommand{\arraystretch}{1.2}
    \caption{Radio properties of the emission components of J1221$-$1354 derived from spectral analysis.}
    \begin{tabular}{l c c c c c c c c c c}
        \hline
         & (1) & (2) & (3) & (4) & (5) & (6) & (7) & (8) & (9) \\
        Component & $\overline{\alpha}$ & $\alpha_{\rm VLASS}$ & $I_{\rm 0.9 GHz}$ & $S_{\rm 0.9 GHz}$ & $B_{eq}$ & $t_{\rm JP}$ & $\chi^2_{red,\rm JP}$ & $t_{\rm CI}$ & $\chi^2_{red,\rm CI}$ \\
         & & & (mJy\,beam$^{-1}$) & (mJy) & (\textmu G) & (Myr) & & (Myr) & \\ \hline \hline
        JNT & $-$1.0 & -- & 0.6 & 10 & 4.4 & $20.3\pm5.0$ & 1.65 & $37.0^{+8.4}_{-6.1}$ & 6.7 \\
        JN & $-$1.8 & -- & 0.6 & 21 & 9.3 & $30.4\pm2.4$ & 2.09 & -- &  -- \\
        NW & $-$1.9 & -- & 0.7 & 10 & 10.2 &  $32.8\pm3.7$ & 2.09 & -- & -- \\
        JS & $-$2.0 & -- & 1.1 & 24 & 11.3 & $34.1\pm2.1$ & 2.09 & -- & -- \\ \hline
        C$^\dag$ & $-$0.5 & $-$0.6 & 196 & -- & -- & -- & -- & -- & --  \\ 
        SW$^\dag$ & -- & $-$2.3 & 8.8$^{\ddag}$ & -- & -- & -- & -- & -- & -- \\
        SE$^\dag$ & -- & $\le-1.9$ & 3.2$^{\ddag}$ & -- & -- & -- & -- & -- & -- \\ \hline
    \end{tabular}
    \tablefoot{Compact components are indicated by $\dag$. Column 1: the 0.9\,GHz intensity weighted average spectral index. Column 2: the spectral index between MALS 1.6\,GHz and VLASS 3\,GHz. Column 3: the mean surface brightness (for diffuse components) and peak flux density (for compact components) at 0.9\,GHz. For SW and SE, peak flux densities at 1.6\,GHz are provided ($\ddag$). Column 4: the integrated flux density at 0.9\,GHz. Column 5: the derived magnetic field strength. Columns 6 and 7: the spectral age and chi-square for the JP model, respectively. Columns 8 and 9: the spectral age and chi-square for the CI model.}
    \label{tab:specage}
\end{table*}

To quantify the spectrum and derive physical quantities such as the spectral age, we use the Broadband Radio Astronomy Tools \citep[BRATS,][]{Harwood2013,Harwood2015}. We smooth all SPW images to a common resolution of $12''.5\times12''.5$, excluding images from frequency ranges corresponding to four SPWs affected by RFI: SPW 5 (1137-1197\,MHz), 6 (1190-1250\,MHz), 7 (1244–1304\,MHz) and 12 (1511–1572\,MHz). The spectral analysis is performed on the remaining 11 SPW images. The resulting spectral index map (Fig.~\ref{fig:radio_continuum}, \textit{bottom-left} panel) obtained using a local detection threshold of $3\sigma$, confirms the steep spectral index ($\alpha < -1.5$) of JS and JN, a shallower spectrum for C, and notably a shallower spectrum at the northern edge and tip (hotspot) of JN (JNT, $\alpha\sim$ -1.0). We determine an injection index of $\alpha = -0.82$ using the routine \texttt{findinject} in BRATS. This value is determined for the entire source, as is commonly done in the literature \citep[e.g.][]{Harwood2013,Harwood2015}. Because the injection index and spectral age are strongly covariant over the limited frequency range available, the per-component fits are not well constrained.

To estimate the magnetic fields in the different components of the source, we follow \citet{Govoni2004}, assuming $k= U_{pr}/U_{el} = 1$, the ratio of energy density in relativistic protons to that in radiating electrons. We account for the fact that the difference in surface brightness in the jets is likely due to relativistic boosting, setting the surface brightness at 0.9\,GHz to $I_{\rm 0.9 GHz} = 1.0\,\mathrm{mJy\,beam}^{-1}$ for all components. Even if the emission is not boosted, this is an adequate approximation for the observed surface brightness of the jet components (Tab.~\ref{tab:specage}), given the weak dependence of the derived magnetic field on surface brightness.
The least constrained input is the thickness of the jets $d$, as the brightest components of the jet appear unresolved. Given that information, we assume a thickness of the jets of $d = 2$ kpc, i.e. $\sim$1$\arcsec$. However, because the magnetic field strength scales as $d^{-2/7}$, this assumption has little effect on the derived magnetic field strengths.
The resulting equipartition magnetic field (B$_{eq}$) strengths are given in Tab.~\ref{tab:specage}, showing that all jet components save JNT have similar magnetic field strengths. Thus, for further spectral analysis, we treat the JNT component separately to the rest of the jet system. For JNT we adopt a magnetic field value of 0.44 nT, whereas for the rest of the jet system we adopt a single value of 1 nT, consistent with their individually derived B$_{eq}$. Note that BRATS includes inverse-Compton losses on the CMB \citep{Stroe2014}, which at $z$ = 0.1027 correspond to an equivalent magnetic field of $B_{\rm CMB}$ $\sim$ 0.38\,nT.
This is comparable to the adopted field for JNT (0.44\,nT), implying that the derived age for this component is more sensitive to the assumed field strength than those of the other jet components.

Fitting Jaffe-Perola \citep[JP,][]{Jaffe1973} ageing models yields the spectral age map shown in Fig.~\ref{fig:radio_continuum} (\textit{bottom-right} panel), with the results for each component presented in Tab.~\ref{tab:specage}. Similar results are obtained with a Tribble model \citep{Tribble1993}. The mean ages of JS, JN, and the NW component broadly agree, with ages ranging from 30-35\,Myr. There is no clear age gradient from the core, likely due to low S/N and/or the complex geometry of a precessing jet.

Fitting the JP ageing models also gives a younger age for JNT at around 20\,Myr, suggesting rejuvenation of the electron population in this region. Given the evidence for ongoing injection at this location, we also fit a continuous injection (CI) model \citep{Pacholczyk1970} to JNT, using the integrated flux densities of the region as a function of frequency. We do not apply the CI model elsewhere, as there is no comparable evidence for ongoing injection in the other jet components. The resulting spectral age, shown in Tab.~\ref{tab:specage}, is more in line with the spectral ages of the other components. This suggests that this part of the jet could be as old or older than the rest of the jet system, but is continuously replenished with fresh particles. However, the CI model is disfavoured over the single-injection JP model based on the $\chi^2$ values (Tab.~\ref{tab:specage}). We therefore retain the JP-derived young age for JNT as the preferred interpretation, consistent with the reorientation scenario discussed in Section~\ref{sec:discussion}.

\section{Discussion}
\label{sec:discussion}

The morphology of J1221$-$1354 shows the inversion symmetry typical of precessing AGN. However, several morphological peculiarities require further discussion.

\subsection{Peculiarities in the jet morphology}

As noted in Sect.~\ref{sec:radio}, JS is significantly brighter than JN, while JNT extends $2'$ from C, whereas JS reaches only $1\farcm1$. This combination is not straightforward to explain. Precession is required by the inversion-symmetric morphology, but is intrinsically symmetric; both jets sweep the same cone and cannot render one side brighter or longer than the other. Relativistic beaming of an approaching JS would enhance its surface brightness, but would also make it appear longer than the receding jet, opposite to what is observed. The anti-correlation between brightness and length therefore requires an asymmetry in the medium into which the jets propagate, with beaming on top of it modulating the observed brightness ratio. Such environmental asymmetry is invoked in hybrid-morphology sources, where jets from a single engine exhibit different morphologies on the two sides \citep[e.g.][]{GopalKrishna2000, Harwood2020}. With all this in mind for J1221$-$1354, both the higher velocity and younger plasma of JN would be the result of a `jet reorientation event', as described in \citet{Nolting2023}, which can occur when the jet re-enters a cavity of previously deposited plasma. This causes the northern jet to accelerate until it encounters the higher-density surrounding medium, where re-acceleration naturally explains the much younger electron population at the northern edge (Sect.~\ref{sec:specage}). The brighter and shorter morphology of JS then requires that it propagates into a denser medium, an asymmetry that perhaps predates the jets themselves.

Beyond the main jets, faint diffuse emission can be seen surrounding the source. This diffuse emission is too faint to determine spectral index or age on a pixel-by-pixel basis. However, we can characterize the spectrum of this diffuse emission using the same set of smoothed SPW images that we used for the spectral analysis with \textsc{BRATS}. As before, SPWs 5, 6, 7, and 12 are excluded. The flux density in each spectral window is obtained by integrating over a rectangular region covering the source and its immediate environment, excluding all pixels with a measured spectral index (Fig.~\ref{fig:radio_continuum}, \textit{bottom-left}). Integrating the flux over the remaining area yields significant ($>5\sigma$) detections of the faint emission in each SPW. Fitting a power law spectrum yields $\alpha=-3.06\pm0.14$ for this emission. Such a steep spectrum suggests this emission is even older than the main jets, supporting the notion that more than one precession cycle has already occurred, as required for a reorientation event. 

As for SE, it is orthogonal to the main jet axis and its emission does not coincide with any optical galaxy (Fig.~\ref{fig:radio_continuum}). It may trace plasma deposited by the same jet system during an earlier phase of the precession cycle, when the jet axis pointed in this direction. Alternatively, it might represent another jet, either from the companion galaxy or a second SMBH in the core.  We note here the NW component, which could represent the counter-jet of SE, although it is much fainter and more diffuse. We examine these possibilities in the context of precession model below.

\subsection{Origin of precession}
\label{sec:precession}

\begin{figure}
    \centering
    \includegraphics[width=\hsize]{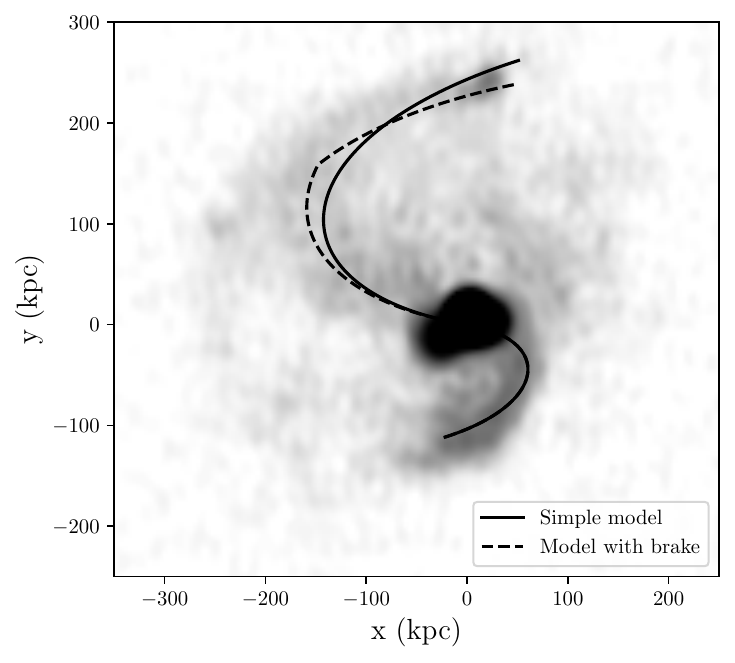}
    \caption{
    0.9\,GHz radio emission of J1221$-$1354 with ballistic precession models overlaid. The simple precession model (solid line, App.~\ref{app:basic}) yields a precession period of $P = 290 \pm 30$\,Myr, and requires the receding JN to have a higher velocity than the approaching JS. To account for the younger plasma age at the northern edge of JN, we add a `brake' to the basic model (dashed line, App.~\ref{sec:brake}), slowing down plasma beyond a specific braking distance. The best-fit braking distance for this model is $d_{brake} = 1000\pm100$\,kpc (see also Section~\ref{sec:precession}).}
    \label{fig:models}
\end{figure}

We model the jets of J1221$-$1354 with a simple ballistic precession model (details in App.~\ref{app:jet_model}), with best-fit models shown in Fig.~\ref{fig:models}. These best-fit models have a precession period of $P=290\pm30$ Myr, and a jet dynamical age of $\sim$30\,Myr, most strongly constrained by the observed flux density ratio and size difference between JS and JN. This dynamical age obtained from the fitting is consistent with the mean spectral age for both jets (Sect.~\ref{sec:specage}). 
Evolving the model further backwards or forwards shows that the jet angle passes through the SE-NW axis at other phases of the precession cycle, so these features may trace plasma deposited by the same jet system during earlier episodes. This is consistent with their steep spectral indices, indicating aged plasma rather than the flat spectrum expected of an independently active core (Tab.~\ref{tab:specage}, \textit{right}). While the observed morphology would be difficult to explain, especially with a jet continuously depositing plasma along the full track, it is still worth noting.

Note that while this model yields consistency between the dynamic and spectral ages of the jets, it produces viewing angles of ${\sim}25^{\circ}$, shallower than typically expected for a narrow line system. While the obscuring torus need not be aligned with the precession axis, this adds to the peculiarity of the source. The shallow viewing angle in the model also means that the jet extends much further out in the $z$-direction than in the image plane. This causes the relatively large `braking distance' of 1000\,kpc for the brake model shown in Fig.~\ref{fig:models}. While it is possible that the viewing angle is larger than obtained from the modelling, this would require larger jet speeds to reproduce the observed brightness asymmetry between JS and JN, but would also yield smaller dynamical ages. Alternatively, if a part of this brightness asymmetry arises from the denser environment of JS (Sect. 4.1) rather than from Doppler boosting, a larger viewing angle would be permitted without increasing the jet speeds, easing the tension with the narrow-line classification.

Given that more than one precession cycle has likely occurred, geodetic precession caused by SMBHB interaction is a plausible origin. Assuming precession is driven by a binary companion of mass $m$ to the central SMBH, the precession period ($P$) is tied to the black hole mass ($M$) and binary separation ($r$) \citep{Begelman1980,Roos1987}:
\begin{equation}
    P \sim \frac{M}{m}\left[\frac{r}{\mathrm{pc}}\right]^{5/2}\left[\frac{M}{10^8 \mathrm{M_{\odot}}}\right]^{-3/2} \mathrm{Gyr}.
\end{equation}
Using $m_V = 17$ \citep{Veron-Cetty2010} and neglecting K-corrections and extinction, we derive $L_V = 3.25\times10^{10}\,L_\odot$. Applying the $M-L$ relation of \citet{Gultekin2009} then gives $M = (2.5\pm2.2)\times10^8\,M_\odot$, where the uncertainty reflects the intrinsic scatter in the relation. With $P = 290\pm30$\,Myr, this yields a binary separation of $r = 1.06\pm0.56$\,pc for $M/m = 1$, representing an upper limit given that $m < M$. 

We now consider observational signatures that could reveal a binary directly.
We searched for double-peaked emission lines and multiple VLBI components associated with C, both of which may provide evidence for a SMBHB \citep{Nandi2021}. While we do not observe double-peaked emission lines in the spectrum of J1221$-$1354 (Fig.~\ref{fig:spectrum}), this does not preclude the presence of a SMBHB. This is because the typical peak separation of $\sim$400\,km\,s$^{-1}$ \citep{Nandi2021} may not be resolved at the spectral resolution and S/N of the 6dFGS spectrum. 
At radio wavelengths, as noted in Sect.~\ref{sec:data}, the VLBI images show a faint component visible north-east of C, at a distance of $3.2\ \mathrm{mas} = 6.5\ \mathrm{pc}$.
This is too far to be a companion driving precession with the inferred period \citep[cf.][]{Nandi2021}. 
The predicted separation of $\lesssim$1.1\,pc corresponds to 0.56\,mas at this redshift, below the minor axis of the X-band beam (1.2\,mas), so a binary at this separation would not be resolved in the available VLBI data.
Either a different mechanism drives the precession, or the VLBI component represents the base of JN.
%

Several observations would test these scenarios. Longer-term VLBI monitoring could establish whether the component 6.5\,pc from the core shows proper motion, as expected if it is a jet feature. The VLASS data, taken at three epochs in 2019, 2022, and 2024, show a $\sim$10 per cent increase in flux density between 2022 and 2024. However, this is comparable to the flux-scale accuracy of VLASS, so dedicated monitoring would be required to establish whether the core is genuinely variable. Polarimetry would test the environmental asymmetry as the denser gas around JS would produce higher Faraday rotation and stronger depolarization \citep[][]{Saikia2003}, opposite to the Laing-Garrington effect expected if JS is the approaching jet.

The nature of the nearby galaxy also remains open. Its SPHEREx redshift is inconsistent with that of the host, but the spectrum is likely contaminated at a separation of 6$\arcsec$ (App.~\ref{app:spherex}), while its optical colours and galaxy type suggest association. A similar configuration is seen in PKS\,B2300$-$189, another known precessing AGN \citep{Hunstead1984,Misra2025}, where the companion direction is also tangential to the jet axis. How such a configuration would drive precession is unclear, and the optical images show no clear evidence for interaction. Deeper spectroscopy would settle whether the two galaxies are physically associated, and hence whether the companion is a plausible participant in the history of this system.

\section{Conclusion}\label{sec:conclusion}

We present J1221$-$1354, a precessing AGN discovered in MALS, with an `S'-shaped morphology, revealed by the excellent surface brightness sensitivity of MeerKAT. The northern jet (JN) is longer and faster than the southern jet (JS), with its tip (JNT) showing a significantly younger spectral age. Both these features are consistent with a jet reorientation event \citep{Nolting2023}, in which the jet re-enters a cavity of previously deposited plasma in an asymmetric ambient medium. A simple ballistic model yields $P = 290\pm30$\,Myr, consistent with geodetic precession in a SMBHB with a binary separation of less than 1.1\,pc. This separation is not resolved with available VLBI data. 
Long-term VLBI monitoring, polarimetry, and high-resolution radio imaging would clarify the nature of the radio core, the role of the ambient medium, and the origin of the diffuse emission associated with the source. Deep radio continuum surveys in the near future will uncover many more such systems, helping to strengthen the connection between jet precession and SMBHBs, and the latter's implications for the gravitational-wave background detected with pulsar timing arrays \citep[e.g.,][]{Agazie2023,EPTACollaboration2023}.

\begin{acknowledgements}
We thank the anonymous referee for their insightful comments. The authors wish to thank Gunther Witzel, Andrei Lobanov, Vijay Mahatma, and Roland Timmerman for many helpful and stimulating discussions.
The MeerKAT telescope is operated by the South African Radio Astronomy Observatory, which is a facility of the National Research Foundation, an agency of the Department of Science and Innovation. The MeerKAT data were processed using the MALS computing facility at IUCAA (\url{https://mals.iucaa.in/releases}).
\end{acknowledgements}

\bibliographystyle{bibtex/aa}
\bibliography{bibtex/MALS-AGN}

\begin{appendix}

\section{Optical spectrum}
\label{app:6dfspec}
%
\begin{figure*}[h]
    \centering
    \includegraphics[width=1.0\linewidth]{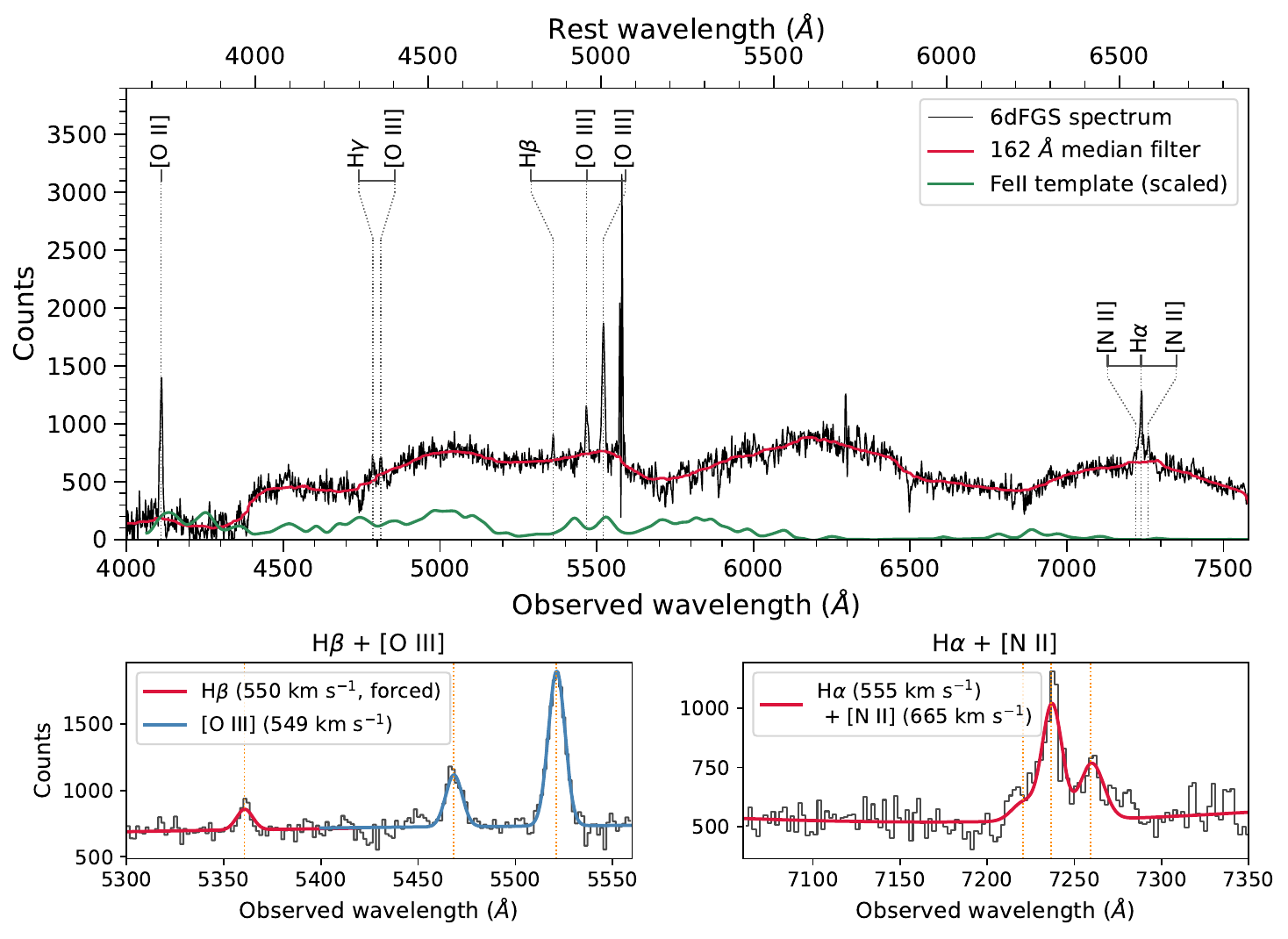}
    \renewcommand{\thefigure}{A.1.}
    \caption{\textit{Top:} 6dFGS optical spectrum of J1221$-$1354 (black), with the underlying continuum represented by an empirical 162\,\AA\ kernel (red). Several emission lines are prominently visible, as well as the 5577\,\AA\ skyline. Fe\,\textsc{ii} template of \citet{Boroson1992}, redshifted and scaled for visibility is also shown (green; ). \textit{Bottom:} Local fits to the H$\beta$+[\ion{O}{iii}] (\textit{left}) and H$\alpha$+[\ion{N}{ii}] (\textit{right}), with fitted FWHM given in the legend (see Appendix~\ref{app:6dfspec} for details).}
    \label{fig:spectrum}
\end{figure*}

Figure~\ref{fig:spectrum} shows the optical spectrum of J1221$-$1354 as observed by 6dFGS \citep{Jones2004}, consisting of independent V-arm (3924–5598\,\AA) and R-arm (5318–7582\,\AA) exposures with different gratings. This target is part of the Type~2 AGN sample of \citet{Suresh2025}, who catalogue FWHM of H$\alpha$ = 556.4 $\pm$ 49.5\,km\,s$^{-1}$, H$\beta$ = 599.8 $\pm$ 35.6\,km\,s$^{-1}$ and [\ion{O}{iii}] = 614.2 $\pm$ 48.0\,km\,s$^{-1}$.  Their estimated ratios of $\log([\ion{O}{iii}]/\mathrm{H}\beta)$ = 1.00 and $\log([\ion{N}{ii}]\,\lambda6583/\mathrm{H}\alpha)$ = $-$0.27 classify this source as a Type-2 AGN. 
Their reported FWHM of [\ion{N}{ii}] = 706.0\,km\,s$^{-1}$ carries a formal uncertainty of exactly zero. It is among the largest values anywhere in their catalogue, which shows a sharp ceiling at $\sim$700\,km\,s $^{-1}$, consistent with the 700\,km\,s$^{-1}$ upper bound imposed on this parameter in their fitting procedure. It is therefore not a measurement of the width of [\ion{N}{ii}] in this source. 
In general, 6dFGS spectra are not flux calibrated, and \citet{Suresh2025} note that $\sim$12\% of the $\sim$5400 spectra they reject on visual inspection show wavy or curved continua attributable to fibre transmission or data reduction issues. They mitigate these issues by fitting the V and R arms separately. This, together with the unconstrained [\ion{N}{ii}] width noted above, motivates a closer look at the spectrum of J1221$-$1354 presented below.

As shown in Fig.~\ref{fig:spectrum}, the continuum displays low-level undulations on scales of several hundred \AA, well in excess of the widths of the emission lines of interest. An empirically adopted running median filter with a 162\,\AA\ kernel representing a reasonable underlying continuum is overlaid on the observed spectrum.
However, a formal periodogram analysis of the continuum residuals shows only marginal power near this scale in the V-arm, with no corresponding feature in the independently observed R-arm spectrum. Crucially, these wobbles exhibit scales far broader than the emission-line regions of interest and both members of each emission-line ratio used in the BPT analysis fall within a single arm, so are unaffected by any cross-arm miscalibration. 
Fe\,\textsc{ii} multiplets from \citet[][]{Boroson1992}, which can contaminate the H$\beta$ region in broad-line AGN, are also shown for reference in Fig.~\ref{fig:spectrum}. As shown below, no broad-line component is detected in this source, so no Fe\,\textsc{ii} contribution is expected, and indeed the predicted contamination at this scaling is negligible relative to the noise.
Therefore, rather than adopting a global continuum model \citep[cf.][]{Hviding2022}, we fit each line locally with a low-order
polynomial continuum.

Fixing the doublet ratios by atomic physics, we find that H$\alpha$ (FWHM $= 555 \pm 55$\,km\,s$^{-1}$) and [\ion{O}{iii}] (FWHM $= 549 \pm 21$\,km\,s$^{-1}$) independently give consistent widths, with no evidence for an additional broad or velocity-offset (outflow) component in either line \citep[cf.][]{Deka2026}. H$\beta$, however, is significantly ($\sim$3$\sigma$) narrower (FWHM $= 291 \pm 66$\,km\,s$^{-1}$; see Fig.~\ref{fig:spectrum}). Tying H$\alpha$ and H$\beta$ to a common width, as is nominally done \citep[][]{Hviding2022, Deka2026}, provides a compromised FWHM $= 488 \pm 45$\,km\,s$^{-1}$ for the Balmer lines, but at the expense of broadening [\ion{N}{ii}] from $665 \pm 118$ to $740 \pm 129$\,km\,s$^{-1}$. 
We attribute the H$\beta$ anomaly to unmodelled underlying stellar absorption expected in NLAGN. We therefore adopt the NLR width ($550$\,km\,s$^{-1}$) established independently by H$\alpha$ and [\ion{O}{iii}]  for H$\beta$, and treat the resultant flux as a lower limit. This gives $\log([\textrm{N}\,\textsc{ii}]/\mathrm{H}\alpha) = -0.23 \pm 0.09$ and $\log([\textrm{O}\,\textsc{iii}]/\mathrm{H}\beta) \leq +0.89$. 
For a 1--3\,\AA\ stellar absorption, spanning the range expected for evolved stellar populations, the source remains $0.1$--$0.4$\,dex above the maximum-starburst line of \citet{Kewley2001}, consistent with the AGN classification of \citet{Suresh2025} for this source.  The NLAGN classification is therefore preserved for any plausible absorption correction, although the margin above the maximum-starburst line is modest.

\section{Redshift estimation using SPHEREx}\label{app:spherex}

\begin{table}[t]
    \centering
    \caption{2MASS sources within 2\farcm5 of J1221$-$1354 for which SPHEREx spectra were recovered and fitted to obtain photometric redshifts.}
    \renewcommand{\arraystretch}{1.2}
    \begin{tabular}{c c c c}
        \hline
        (1) & (2) & (3) & (4) \\
        2MASS ID & $z$ & $\theta$ & $d$ \\ 
                 &     &  ($''$)    & (kpc) \\ \hline
        J12213110-1354016 & $0.105\pm0.001$ & 0 & 0 \\
        J12213138-1354065 & $0.058\pm0.002$ & 6 & 13 \\
        J12213079-1353395 & $0.008^{+0.002}_{-0.001}$ & 22 & -- \\
        J12212891-1353503 & $0.015^{+0.007}_{-0.004}$ & 34 & -- \\
        J12213451-1354053 & $0.098\pm0.005$ & 50 & 97 \\
        J12213456-1353536 & $0.090^{+0.005}_{-0.004}$ & 51 & 100 \\
        J12212613-1354058 & $0.097\pm0.008$ & 72 & 141 \\
        J12212589-1354223 & $0.087\pm0.002$ & 79 & -- \\
        J12212733-1353005 & $0.102^{+0.005}_{-0.004}$ & 82 & 160 \\
        J12212930-1355392 & $0.010^{+0.003}_{-0.001}$ & 101 & -- \\
        J12212638-1355274 & $0.005^{+0.001}_{-0.0004}$ & 110 & -- \\
        J12213827-1353107 & $0.007^{+0.002}_{-0.001}$ & 116 & -- \\
        J12213076-1352044 & $0.030\pm0.003$ & 117 & -- \\
        J12213295-1356092 & $0.092\pm0.004$ & 130 & 254 \\
        J12213992-1353133 & $0.098\pm0.002$ & 137 & 268 \\
        J12212143-1354543 & $0.106\pm0.002$ & 150 & 293 \\
        J12212467-1351519 & $0.008\pm0.001$ & 159 & -- \\
        \hline
    \end{tabular}
    \tablefoot{Column 1: 2MASS identifier. Column 2: the best-fit redshift posterior from \textsc{Bagpipes}. Column 3: the angular distance to the optical host of J1221$-$1354. Column 4: the project physical distance to the host, calculated assuming the host redshift of $z$ = 0.1027. This is only listed for the sources whose redshifts are consistent with the host redshift within 3\textsigma, except for the potential companion galaxy, J12213138$-$1354065.}
    \label{tab:sources}
\end{table}


We select all 2MASS sources within $2\farcm5$ of J1221$-$1354 (see Tab.~\ref{tab:sources}). We retrieve the SPHEREx spectra for these, and fit them using Bayesian Analysis of Galaxies for Physical Inference and Parameter EStimation \citep[\textsc{Bagpipes},][]{Carnall2018}. While the chosen search radius does not capture the full extent of a galaxy group or cluster \citep[e.g.][]{Tully2015}, it covers the extent of the radio emission, and retrieves a reasonably small sample of sources to perform spectral modelling on. As the spectra do not have the spectral resolution to identify individual lines, redshift estimation relies on large-scale spectral features. At low redshifts, the primary feature with which redshift can be determined in the wavelength range probed by SPHEREx is the 1.6\,$\mu$m stellar bump. Optimising fit parameters for elliptical galaxies, we fit an exponential star-formation history with total stellar mass ($M$) metallicity ($Z$) time since beginning of star formation ($t$) and decay parameter ($\tau$). Additionally, we fit for dust attenuation ($A_V$) and most importantly, redshift ($z$). While elliptical galaxy priors are appropriate for the host and the morphologically similar galaxies identified in DESI images, the 1.6\,$\mu$m stellar bump is most prominent in evolved stellar populations, and these priors may not be ideal for all sources in the field. The resulting redshifts of the nearby bright sources are listed in Tab.~\ref{tab:sources}. The best-fit redshift of J1221$-$1354 is consistent with the spectroscopic redshift within 3\textsigma. Using this significance as a guideline, of the 16 remaining sources, seven redshifts coincide with the host redshift within 3\textsigma, suggesting a moderately over-dense environment. Their calculated physical distances to the host at $z=0.1027$ are also included in Tab.~\ref{tab:sources}.

\section{Modelling of the jets}\label{app:jet_model}

Jet morphologies of precessing AGN can be modelled with a simple ballistics simulation \citep{Hunstead1984,Condon1984a}. In this model, `plasmoids' are ejected in opposite directions from the core with velocity $v=\beta c$. The direction at which they are ejected will evolve according to a precession model, where $\phi$ describes a fixed angle from the precession axis, about which the axis of ejection precesses with precession period $P$. Defining a coordinate system in which the precession axis is in the z-direction, the precession changes the azimuthal angle $\eta = \eta_0 + 2\pi t_e/P$ in which the plasmoids are ejected at each time $t_e$. The position of all plasmoids in this system at a given time $t$ is given by
\begin{align}
    \vec{x}' &= \beta c (\vec{t}-t_e)\sin\phi\cos\eta, \nonumber \\ 
    \vec{y}' &= \beta c (\vec{t}-t_e)\sin\phi\sin\eta, \\ 
    \vec{z}' &= \beta c (\vec{t}-t_e)\cos\phi. \nonumber
\end{align}
It is convenient to transform to a coordinate system where the $z$-axis points towards the observer, aligning the $x,y$-plane with the image plane. This is described by the inclination angle ($i$) of the system, which transforms the coordinates as
\begin{align}\label{eq:final_coords} 
    \vec{x} &= \beta c (\vec{t}-t_e)\sin\phi\cos\eta, \nonumber \\ 
    \vec{y} &= \beta c (\vec{t}-t_e)(\sin\phi\sin\eta\cos i + \cos\phi\sin i), \\ 
    \vec{z} &= \beta c (\vec{t}-t_e)(\cos\phi\cos i - \sin\phi\sin\eta\sin i). \nonumber
\end{align}
To account for relativistic effects, we define the line-of-sight velocity of the plasmoids, $\beta_z = \beta(\cos\phi\cos i - \sin\phi\sin\eta\sin i)$. 
The observed brightness of the plasmoids is
\begin{equation}
    S_{obs} = S \left(\frac{\sqrt{1-\beta^2}}{1-\beta_z}\right)^{k-\alpha},
\end{equation}
where $k=2$ for a continuous jet, and $\alpha$ is the spectral index. The apparent length of the jets is also affected as a consequence of light from the receding jet reaching the observer later than light from the approaching jet. The observed radiation time, $t_r$, is 
\begin{equation}
    \vec{t}_r = \vec{t} - \frac{\vec{z}}{c} = \vec{t} - \beta_z (\vec{t}-t_e).
\end{equation}
This quantity replaces $\vec{t}$ in Eq.~\ref{eq:final_coords}. 
An exhaustive parameter search, e.g., via Monte Carlo sampling, is impractical. We instead present models that fit the data to a satisfactory degree.

\subsection{Basic model}
\label{app:basic}

With the observed shape and spectral ages, we can make a few prior assumptions. While the spectral age suggests some plasma is as old as 35\,Myr, a stronger constraint comes from the difference in brightness between the two jets, JS and JN. Therefore, models that reproduce the surface brightness ratio between the two jets, which is $S_{JS}/S_{JN} \approx 2.6$, are favoured. From the difference in length between JS and JN, it is clear that this cannot be modelled by jets moving at the same speed $\beta$. Therefore, we assume $\beta_{\rm JN} > \beta_{\rm JS}$. Since the ends of the JN and JS point straight north and south respectively, we set $\eta_0=\pi/2$.

As an automated parameter search using a least squares or likelihood optimisation routine is impractical, the model is optimised by visual comparison between the model projected onto the image plane and the observed radio emission. To supplement this, each plasmoid is injected as a point source into a model image matching the dimensions of the radio data, and convolved with the beam. This model image is subtracted from the original image, taking the sum-of-squares of the residual as a goodness-of-fit metric. Parameter uncertainties are estimated by varying each parameter incrementally while keeping all others fixed, until the visual fit is no longer adequate and/or the residual sum-of-squares has appreciably increased.

The best-fit model has $i=30^{\circ}\pm1^{\circ}$, $\phi=40^{\circ}\pm1^{\circ}$, $P=290\pm30$ Myr, $\beta_\mathrm{JS}=0.062\pm0.003$, and $\beta_\mathrm{JN}=0.19\pm0.01$, and is shown in Fig.~\ref{fig:models}. The precession period is naturally degenerate with the respective speeds of the jets, and is primarily constrained by the observed jet brightness ratio, the spectral age, and the length of the jets. While this result reproduces the jet brightness ratios, it requires the visible jets to have an age of around 32\,Myr, broadly consistent with the observed spectral age. Given that the spectral age depends on additional assumptions such as magnetic field and injection index, we consider the brightness ratio a more direct and reliable constraint than the spectral ages.

\subsection{Model components with brake}
\label{sec:brake}

While the basic model specified in App.~\ref{app:basic} matches the observed morphology closely, it does not account for the younger spectral age of the northern end of the jet. We therefore add a `brake' at a certain distance from the core, $d_{brake}$. This represents interaction with the surrounding medium, beyond which the jet has a reduced velocity, potentially creating a shock which rejuvenates the electron population, accounting for the younger plasma age in that region. We also add a post-brake velocity $\beta_{brake}$, which the plasmoids have after reaching the brake distance. Initialising all the other parameters with the `best-fit' values from Sect.~\ref{app:basic}, we find an increased $\beta_{JN} = 0.22\pm0.03$, along with $d_{brake} = 1000\pm100$ kpc and $\beta_{brake}=0.12\pm0.02$. Note that $d_{brake}$ is large in projection, but consistent with the jet's three-dimensional extent given the shallow viewing angle (Sect.~\ref{sec:precession}). As shown in Fig.~\ref{fig:models}, the resulting jet model closely resembles the observed morphology. This model is also consistent with a `jet reorientation event', as described in MHD simulations by \citet{Nolting2023}.

\end{appendix}

\end{document}